# Viscoelastic optical nonlocality of low-loss epsilon-near-zero nanofilms

Domenico de Ceglia[a,1], Michael Scalora[b], Maria A. Vincenti[c], Salvatore Campione[d], Kyle Kelley[e], Evan L. Runnerstrom[e], Jon-Paul Maria[e], Gordon A. Keeler[f], and Ting S. Luk[d,f]

[a] *Department of Information Engineering, University of Padova, Italy*
[b] *US Army AMRDEC, Charles M. Bowden Research Laboratory, Redstone Arsenal (AL), USA*
[c] *Department of Information Engineering, University of Brescia, Italy*
[d] *Sandia National Laboratories, Albuquerque (NM), USA*
[e] *Department of Materials Science, North Carolina State University, Raleigh (NC), USA*
[f] *Center for Integrated Nanotechnologies (CINT), Sandia National Laboratories, Albuquerque (NM), USA*
[1] *Corresponding author:* domenico.deceglia@unipd.it

**Abstract**

Optical nonlocalities are elusive and hardly observable in traditional plasmonic materials like noble and alkali metals. Here we report experimental observation of viscoelastic nonlocalities in the infrared optical response of doped cadmium-oxide, epsilon-near-zero nanofilms. The nonlocality is detectable thanks to the low damping rate of conduction electrons and the virtual absence of interband transitions at infrared wavelengths. We describe the motion of conduction electrons using a hydrodynamic model for a viscoelastic fluid, and find excellent agreement with experimental results. The electrons' *elasticity* blue-shifts the infrared plasmonic resonance associated with the main epsilon-near-zero mode, and triggers the onset of higher-order resonances due to the excitation of electron-pressure modes above the bulk plasma frequency. We also provide evidence of the existence of nonlocal damping, i.e., *viscosity*, in the motion of optically-excited conduction electrons using a combination of spectroscopic ellipsometry data and predictions based on the viscoelastic hydrodynamic model.

**Introduction**

The interaction of light with electrons in motion produces plasmon polaritons, i.e., light-driven collective oscillations of conduction electrons[1]. Usual electromagnetic signatures of plasmon polaritons include subwavelength optical confinement, absorption peaks in angular and frequency spectra, sensitivity to *p*-polarized light, and large electric-field enhancement[2-3]. Owing to these properties, structures that support plasmonic modes improve sensors[4], photodetectors[5], absorbers[6], and thermal management devices[7], and hold the promise to merge electronics and photonics at the nanoscale[8]. Indeed, current nanotechnologies allow fabrication of plasmonic structures with unprecedented quality and precision, with sub-nanometer and, in some cases, atomic-size features. This ever-increasing miniaturization trend has stimulated the growth of new research areas, such as nanoplasmonics[9], nonlinear plasmonics[10] and quantum plasmonics[11]. A fundamental understanding of microscopic events occurring in the bulk and on the surface of these systems, and the ability to account for these phenomena in simple, yet accurate theoretical models, are necessary steps to advance these fields of studies. Important examples of microscopic interactions include quantum, nonlocal (e.g., spatially-dispersive), and nonlinear effects induced by optically-excited charge carriers in nanoplasmonic structures.

So far, all these effects have been primarily studied in nanostructures based on alkali and noble metals. For example, longitudinal plasmons attributed to the hydrodynamic behavior of conduction electrons were observed in thin films of potassium[12] and magnesium[13-14] using electron-spectroscopy techniques, and in silver films[15] via spectrophotometry. In the case of magnesium, the nonlocal hydrodynamic description of conduction electrons has been validated even for few-atom-thick samples[14]. Nonlocal and quantum effects, such as quenching and shifting of plasmonic resonances, have also been found in isolated[16] and coupled metallic nanoparticles with sub-nanometer gaps[17]. The impact of conduction electrons' spill-out from metal surfaces has been investigated with both ab-initio and quantum hydrodynamic models that are generally in good agreement with experimental results[18-20]. Moreover, nonlinear phenomena related to quantum tunneling in metal-insulator-metal systems have been theoretically predicted via first-principle calculations[21] and semiclassical theories[22-24].

In this paper, we report, for the first time to our knowledge, infrared optical nonlocalities due to conduction electrons in indium-doped CdO thin films. Semiconductors such as conducting-metal oxides and transition-metal nitrides are emerging in the context of plasmonic materials as alternatives to noble metals because of their low absorption losses[25], CMOS compatibility, and tunability, thus offering new opportunities for the realization of photonic and electro-optic devices[26] in the infrared range. Among these materials, indium-doped CdO lends itself particularly well to the study of quantum and nonlocal hydrodynamic effects for at least two reasons: (i) the optical properties of doped CdO are dominated by conduction electrons near the plasma frequency, where interband transitions are practically absent, similarly to other doped conducting-metal oxides (e.g., indium tin oxide and aluminum zinc oxide); (ii) absorption losses can be controlled with doping – films with carrier mobility around 500 $cm^2/(V \cdot s)$ and imaginary part of the (local) bulk dielectric constant as low as 0.15 have been measured near the zero-crossing frequency of the real part of the bulk dielectric constant[27].

It is well known that thin planar films support a polaritonic mode near the zero-crossing frequency of the real part of the permittivity, usually referred to as either bulk-plasma frequency or epsilon-near-zero (ENZ) frequency. This mode, also known as ENZ mode and Ferrell-Berreman mode[28-29], is plasmonic in metallic materials, phononic in dielectrics, and can be both *leaky* (radiative) or confined (non-radiative)[30]. Optical excitation of ENZ modes generally requires an electric field component normal to the plane of the film, i.e., *p*-polarized light at oblique incidence. The result is the enhancement of the normal electric-field component inside the film and the formation of an absorption peak in the spectrum[31]. So far, these phenomena have been discussed within the local approximation of the film's dielectric constant, in which the relative permittivity depends only on the frequency, i.e., $\varepsilon_L = \varepsilon'(\omega) + i\varepsilon''(\omega)$. The strength of light-coupling to the ENZ mode, as well as the amount of normal electric-field enhancement, increases with the decay time of the mode, a quantity that is inversely proportional to the imaginary part of the relative permittivity, $\varepsilon''(\omega)$. In metals, $\varepsilon''(\omega)$ is quite large at the zero-crossing frequency of $\varepsilon'(\omega)$, with damping rates worsened by interband transitions. For this reason, research has focused on artificial (effective) ENZ conditions obtained with periodic structures (metamaterials) and waveguides at cutoff. The issues that arise using these approaches are the presence of scattering and the difficulty in miniaturization. On the other hand, an alternative route to observing strong coupling to ENZ modes is provided by films of doped semiconductors, where conduction electrons respond as a Drude-like plasma in the visible and infrared ranges, with $\varepsilon''(\omega)$ values as low as 0.5 for ITO[32] and 0.15 for dysprosium-doped CdO[27]. A number of interesting phenomena have been observed in these systems. For example, complete light absorption has been demonstrated in the Kretschmann configuration[30] (i.e., light is prism-coupled to the film) and using a back mirror[33]. Moreover, thanks to the large electric-field enhancement, the efficiency of nonlinear optical effects is boosted: enhanced third-harmonic generation has been demonstrated in thin indium-tin-oxide[32, 34] ENZ films. Ultrafast nonlinear behavior, with near-unity refractive index changes, has been recently detected in ENZ films of indium-tin-oxide[35], aluminum-zinc-oxide[36] and doped cadmium-oxide[37].

In our experiments, In:CdO planar films with thicknesses in the 10-20 nm range are excited with *p*-polarized light in the epsilon-near-zero (ENZ) regime, i.e., near the zero-crossing frequency of the films (local) dielectric constant. Previous studies have focused on the fundamental properties of ENZ modes within the local approximation for the conduction-electron current density[38-39], summarized by the Ohm's law $\mathbf{J}(\mathbf{r}) = \sigma \mathbf{E}(\mathbf{r})$, where $\mathbf{J}(\mathbf{r})$ and $\mathbf{E}(\mathbf{r})$ are, respectively, the induced current density and the local electric field at the position **r**, and $\sigma$ is the complex, frequency-dependent conductivity. Here we experimentally demonstrate via optical spectroscopy that the phenomenology of ENZ modes in doped CdO nanofilms significantly deviates from the local picture of Ohm's law. We use a mesoscopic classical model based on the hydrodynamic theory of viscoelastic fluids[40-41] to describe the behavior of conduction electrons and find qualitative and quantitative agreement with experimental results. Mesoscopic models, such as those based on the hydrodynamic theories[41-43], the quantum corrected model[22], and the quantum conductivity theory[23], are particularly interesting and useful for their ability to bring quantum statistical and quantum mechanical effects within the framework of classical electrodynamics, with little or no impact on the computational-power requirements. These models represent valid alternatives to purely quantum mechanical treatments – e.g., those based on the density functional theory – that may be significantly limited by current computational power when dealing with problems involving structures with thousands or millions of atoms. The model that we adopt generalizes the Bloch hydrodynamic theory[1] and shows that the nonlocal hydrodynamic interaction of the electron-fluid with *p*-polarized light induces: (i) a significant blue-shift of the dominant ENZ mode; (ii) additional, thickness-dependent resonances above the bulk plasma frequency of the film associated with the elasticity of electrons and higher-order pressure modes; and (iii) nonlocal damping due to the electrons' viscosity that augments absorption losses.

**Results**

Our objective is to detect and quantify nonlocal, viscoelastic effects of conduction electrons in nanofilms of doped semiconductors. Doping engineering of CdO has been recently proven to be a valid strategy to achieve high electron mobility and low-loss ENZ behavior. Here, indium-doped CdO films were deposited on an MgO substrate (thickness 500 μm) via magnetron sputtering. The choice of the MgO substrate is key to keep absorption losses as low as possible in the experiment, so that the spectral features due to the viscoelastic properties of conduction electrons are more detectable. Two samples were fabricated, with thicknesses of 20 and 11 nm, determined by x-ray reflectivity measurement and fitted with an uncertainty less than 2 nm. A root-mean-square (RMS) roughness of 0.7 nm on the CdO films was measured with atomic-force microscopy (AFM). The 20-nm (11-nm) sample was sputtered with an indium-concentration of $\sim 2.7 \times 10^{26}$ m$^{-3}$ ($\sim 3.2 \times 10^{26}$ m$^{-3}$). Reflectivity spectra at oblique incidence were taken with a spectroscopic ellipsometer (J. A. Woollam V-VASE and IR-VASE) in the spectral range of 1.6-10 μm. In the Kretschmann or attenuated-total-reflection (ATR) configuration, the light source was coupled to the samples through a calcium fluoride prism using an index-matching oil between a 90° apex-angle prism and the MgO substrate. Reflectivity spectra for $s$- and $p$-polarized light were retrieved at 45°, above the critical angle of the prism-air interface, $\theta_c = \mathrm{asin}\left(\sqrt{\varepsilon_{\mathrm{Air}} / \varepsilon_{\mathrm{MgO}}}\right)$. In the air-coupling configuration, reflectivity spectra were taken at the same 45° angle. In Fig. 1a we report the normalized reflectivity spectra $R_p/R_s$ for the 20-nm-thick In:CdO film measured in both the ATR and the air-coupling cases, where $R_{p,s}$ indicates the reflectivity for $p/s$ polarizations. A deep resonance, labeled $\lambda_1$, is excited in both illumination conditions and corresponds to a reflection dip around 2.14 μm for $p$-polarized light. The shallower dips at 1.95 and 1.7 μm are additional resonances for $p$-polarized light, labeled $\lambda_3$ and $\lambda_5$, respectively; zooms of the spectra near these features are reported in Fig. 1b (ATR coupling) and Fig. 1c (for air-coupling).

We now clarify the resonance spectrum structure by using the generalized hydrodynamic model outlined by Tokatly and Pankratov[41], which treats conduction electrons as a viscoelastic fluid. We adopt the linear approximation in which light-induced variations of electron density $\delta n$ and electron pressure $\delta P$ are assumed to be perturbations around the equilibrium values $n_0$ and $P_0$, respectively. The model is equivalent to the theory of highly viscous fluids with the conduction electrons response given by the following Navier-Stokes equation:

$$m^* \frac{\partial \mathbf{v}}{\partial t} + \frac{\nabla \delta P + \nabla \cdot \boldsymbol{\pi}}{n_0} + m^* \gamma \mathbf{v} = -e\mathbf{E}, \tag{1}$$

where $\gamma$ is the electron collision frequency, $m^*$ the effective electron mass, $e$ the (positive) elementary charge, $\mathbf{v}$ the velocity field and $\mathbf{E}$ the electric field. Moreover, $\delta P$ is the perturbation of the scalar part of the stress tensor, i.e., $P = P_0 + \delta P$ is the local pressure, and $\pi$ is the trace-less part of the viscoelastic stress tensor. We consider the kinetic pressure of a degenerate Fermi gas at $T = 0$ K, so that $P = \frac{\hbar^2}{5m^*}(3\pi^2)^{2/3} n^{5/3}$; $n = n_0 + \delta n$ is the locally-perturbed electron density. In the second-order approximation of the theory with respect to the ratio $\alpha = \frac{v_F}{L \max\{\omega, \gamma\}}$ ($L$ is the characteristic dimension of the system, 10-20 nm for our nanofilms, and $v_F$ the Fermi velocity,) the dynamics of the scalar ($\delta P$) and trace-less part ($\pi$) of the stress tensor are derived from the following equations[41]:

$$\frac{\partial \delta P}{\partial t} = -K \nabla \cdot \mathbf{v} \tag{2a}$$

$$\frac{\partial \pi_{ij}}{\partial t} + \mu\left(\nabla_i v_j + \nabla_j v_i - \frac{2}{3}\delta_{ij}\nabla \cdot \mathbf{v}\right) = -\gamma \pi_{ij}, \tag{2b}$$

where the coefficients $K = \frac{5}{3}P_0$ and $\mu = P_0$ are the bulk and shear modulus of the electron fluid, and $P_0 = \frac{\hbar^2}{5m^*}(3\pi^2)^{2/3} n_0^{5/3}$. Dissipation is negligible only in the collision-less regime, i.e., $\omega \gg \gamma$ and $\left|\frac{\partial \pi_{ij}}{\partial t}\right| \gg |\gamma \pi_{ij}|$. On the other hand, in the collisional regime, i.e., $\omega \ll \gamma$, dissipative effects are present in the system in the form of viscosity: indeed, eq. 2b predicts $\pi_{ij} = -\frac{\mu}{\gamma}(\nabla_i v_j + \nabla_j v_i - \frac{2}{3}\delta_{ij}\nabla \cdot \mathbf{v})$, which corresponds to the stress tensor of a viscous fluid with viscosity $\eta = -\mu/\gamma = -P_0/\gamma$. In the intermediate scenario in which $\omega$ is approximately 10-times larger than the collision frequency $\gamma$, as is the case for materials like CdO and ITO near their ENZ frequency, it is reasonable to expect some role of viscosity in the optical response. Eqs. 1 and 2, plus the continuity equation

$$\frac{\partial \delta n}{\partial t} = -n_0 \nabla \cdot \mathbf{v} \tag{3}$$

constitute a closed set of equations for the generalized hydrodynamic theory in the second order of $\alpha$. After Fourier-transforming equations 1-3 (note that we maintain the same symbols to indicate frequency- and time-domain quantities, for simplicity), we write the following equations for the current density $\mathbf{J} = -n_0 e \mathbf{v}$:

$$\mathbf{J}_{NL} + \mathbf{J} = \sigma \mathbf{E} \tag{4a}$$

$$\mathbf{J}_{NL} \approx \frac{2v_F^2/5}{\omega(\omega+i\gamma)}\left[\nabla(\nabla\cdot\mathbf{J})\left(1-i\frac{\gamma}{6\omega}\right)+\frac{1}{2}\left(1-i\frac{\gamma}{\omega}\right)\nabla^2\mathbf{J}\right], \tag{4b}$$

where $\mathbf{J}_{NL}$ is the nonlocal portion of the current density, and $v_F = \hbar(3\pi^2 n_0)^{1/3}/m^*$ is the Fermi velocity. The local conductivity in eq. 4a is $\sigma = \dfrac{\varepsilon_0 \omega_p^2}{\gamma - i\omega}$, where $\varepsilon_0$ is the vacuum permittivity, $\omega_p = \sqrt{n_o e^2/(\varepsilon_0 m^*)}$ is the unscreened bulk plasma frequency, $n_o$ is the (constant) equilibrium carrier density, $m^*$ is the effective electron mass, $\gamma$ is the scattering rate due to collisions. In deriving eq. 4b we assumed $\omega$ is about 10-times larger than the collision frequency $\gamma$, so that $\dfrac{\omega^2 - i\omega\gamma}{\omega^2 + \gamma^2} \approx 1 - i\dfrac{\gamma}{\omega}$.

Some observations are now in order relative to the connections of the present model (eq. 4) to other nonlocal hydrodynamic theories. Many classical hydrodynamic models reported in the literature neglect the trace-less part of the stress tensor, $\pi$, so that the nonlocal current lacks the Laplacian term in eq. 4b, and takes the form $\mathbf{J}_{NL} = \dfrac{\beta^2}{\omega(\omega+i\gamma)}\nabla(\nabla\cdot\mathbf{J})$. For example, the Bloch hydrodynamic theory can be directly derived from the model presented here when $\pi=0$, which leads to a purely real value of $\beta^2 = v_F^2/3$, and any dissipation related to nonlocal damping is ignored. A complex value of $\beta^2$ was derived by Halevi in[44], in which the collision-less longitudinal permittivity based on the Boltzmann equation was corrected with the Mermin approach to take into account the relaxation of charge carriers to the locally-perturbed equilibrium ($n = n_0 + \delta n$). In the generalized hydrodynamic theory proposed by Mortensen et al.[42], dissipation effects are introduced in the form of diffusion currents (Fick's law), and $\beta^2 = 3v_F^2/5 + D\gamma - i\omega D$, where the diffusivity $D$ induces nonlocal damping. Despite some similarities with our approach, in[42] the dissipation term $\omega D$ is proportional to $\omega/\gamma$ and the Laplacian term ($\nabla^2 \mathbf{J}$) is absent as in[44]. In contrast, in the present model dissipation originates from the trace-less stress tensor $\pi$ and is proportional to $\gamma/\omega$ [which is an inverse dependence to what shown in[42], see eq. 4b]. The inclusion of viscoelastic effects has been recently discussed in[45] in the context of the quantum hydrodynamic theory[18-20], a more sophisticated approach that allows spatial variations of the equilibrium density $n_0$ and therefore includes effects such as the spill-out of electrons at interfaces and Friedel oscillations in the bulk of plasmonic nanostructures. Here, these effects are deliberately neglected and the nonlocal constitutive relation of eq. 4 is complemented by the simple boundary condition $\mathbf{n} \cdot \mathbf{J} = 0$, where $\mathbf{n}$ is the unit vector normal to the interface. This is a reasonable assumption in the planar geometries under investigation that extend for 10-20 nm in the longitudinal direction, while perturbations of $n_0$ occur within a few angstroms from the interface.

The optical response of the films is governed by the equation of motion for conduction electrons (eq. 4) plus the usual Helmholtz equation,

$$\nabla \times \nabla \times \mathbf{E} - \left(\frac{\omega}{c}\right)^2 \varepsilon_\infty \mathbf{E} = i\omega\mu_0 \mathbf{J}, \tag{5}$$

where the response of valence and inner-core electrons is summarized by the background (constant) relative permittivity $\varepsilon_\infty$ and $\mu_0$ is the vacuum permeability. Manipulation of eqs. 4 and 5 leads to the following nonlocal Helmholtz equation:

$$\vartheta \nabla \times \nabla \times \mathbf{E} - \left(\frac{\omega}{c}\right)^2 \left[\varepsilon_L + \frac{\varepsilon_\infty(\frac{3}{5} - i\frac{4}{15}\frac{\gamma}{\omega})v_F^2}{\omega(\omega + i\gamma)} \nabla^2\right]\mathbf{E} = \mathbf{0} \tag{6a}$$

$$\vartheta = \left[1 - \frac{v_F^2/5}{\omega(\omega+i\gamma)}\left(\frac{\omega}{c}\right)^2 \varepsilon_\infty (4 - i\frac{7}{3}\frac{\gamma}{\omega}) - (1 - i\frac{\gamma}{\omega})\frac{v_F^2/5}{\omega(\omega+i\gamma)}\nabla \times \nabla \times\right], \tag{6b}$$

where $\varepsilon_L = \varepsilon_\infty + \frac{i\sigma}{\omega\varepsilon_0} = \varepsilon_\infty\left(1 - \frac{\omega_p^2/\varepsilon_\infty}{\omega(\omega+i\gamma)}\right)$ is the relative permittivity in the local approximation.

For plane wave propagation in structures with translational invariance, e.g., the planar films under investigation, we can assume the operator $\vartheta \approx 1$ and transform the Helmholtz equation in the k-space domain by using $\nabla \rightarrow i\mathbf{k}$:

$$-\mathbf{k} \times \mathbf{k} \times \mathbf{E} = \varepsilon_L \mathbf{E} - \frac{\varepsilon_\infty(\frac{3}{5} - i\frac{4}{15}\frac{\gamma}{\omega})v_F^2}{\omega(\omega+i\gamma)}\mathbf{k}(\mathbf{k} \cdot \mathbf{E}). \tag{7}$$

In the Helmholtz decomposition, the electric field is written as the superposition of longitudinal (subscript $\ell$) and transverse (subscript $t$) components $\mathbf{E} = \mathbf{E}_\ell + \mathbf{E}_t$. For longitudinal waves $\mathbf{k} \times \mathbf{E}_\ell = \mathbf{0}$, so that:

$$\left[\varepsilon_\infty - \frac{\omega_p^2 + \varepsilon_\infty(\frac{3}{5} - i\frac{4}{15}\frac{\gamma}{\omega})v_F^2 q^2}{\omega(\omega+i\gamma)}\right]\mathbf{E}_\ell = \mathbf{0}, \tag{8}$$

where $q$ is the longitudinal wavenumber. After neglecting damping for simplicity, the dispersion of longitudinal modes is retrieved as

$$\omega(q) = \sqrt{\omega_p^2/\varepsilon_\infty + \frac{3}{5}v_F^2 q^2}. \tag{9}$$

Extracting a factor $\omega(\omega + i\gamma) - (\frac{3}{5} - i\frac{4}{15}\frac{\gamma}{\omega})v_F^2 q^2$ from eq. 8 yields:

$$\left[ \varepsilon_\infty - \frac{\omega_p^2}{\omega(\omega + i\gamma) - (\frac{3}{5} - i\frac{4}{15}\frac{\gamma}{\omega})v_F^2 q^2} \right] \mathbf{E}_\ell = \mathbf{0}, \tag{10}$$

and the usual form of the longitudinal permittivity for a plasma film is recovered

$$\varepsilon_\ell = \varepsilon_\infty - \frac{\omega_p^2}{\omega(\omega + i\gamma) - (\frac{3}{5} - i\frac{4}{15}\frac{\gamma}{\omega})v_F^2 q^2}. \tag{11}$$

For transverse waves, $\mathbf{k} \cdot \mathbf{E}_t = 0$ and the usual dispersion relation $k_t = \omega/c\sqrt{\varepsilon_t}$ is obtained, with

$$\varepsilon_t = \varepsilon_L = \varepsilon_\infty - \frac{\omega_p^2}{\omega(\omega + i\gamma)}.$$

The ENZ modes may thus be thought of as longitudinal modes of the film, excited by *p*-polarized light (because of the longitudinal field requirement) at the frequencies that yield a vanishing longitudinal permittivity. In the local limit ($q \to 0$), the film displays isotropic relative permittivity, $\varepsilon_\ell = \varepsilon_t = \varepsilon_L$, yielding the single resonance associated with the ENZ mode excited at approximately the screened bulk plasma frequency (or ENZ frequency), $\omega_0 = \omega_p/\sqrt{\varepsilon_\infty}$. In the nonlocal picture using the viscoelastic model, the ENZ modes follow the dispersion relation in eq. 8 and occur at the discrete frequencies $\omega_m = \sqrt{\omega_p^2/\varepsilon_\infty + \frac{3}{5}v_F^2 q_m^2}$, where $q_m = m\pi/d$, $m = 1,3,5,...$ is an odd integer corresponding to the mode order and $d$ is the film thickness. The dips in the measured reflection spectra are predicted to correspond to the resonance wavelengths $\lambda_m = 2\pi c/\omega_m$. All resonances collapse to the local ENZ wavelength $\lambda_0 = 2\pi c\sqrt{\varepsilon_\infty}/\omega_p$ in the local picture, which is a valid approximation for large film thicknesses *d*, i.e., $q_m \to 0$. Reflection spectra are calculated by solving equations 4 and 5 using a nonlocal transfer-matrix-method[46]. We retain only second-order spatial derivatives with respect to the longitudinal direction (*z*-axis) in eq. 4, so that the nonlocal portion of the current density reduces to

$$\mathbf{J}_{NL} \approx \frac{\beta_{VE}^2}{\omega(\omega+i\gamma)} \nabla_z^2 J_z \hat{\mathbf{z}}, \tag{12}$$

where $\beta_{VE} = \sqrt{\frac{3}{5} - i\frac{\gamma_{VE}}{\omega}\frac{4}{15}} v_F$ is the complex viscoelastic, nonlocal coefficient and $\gamma_{VE}$ is the viscoelastic damping rate. While according to the viscoelastic theory $\gamma_{VE} = \gamma$, in what follows we will relax this requirement and introduce a factor $\alpha_{VE} = \gamma_{VE}/\gamma$ in order to optimize the fit of the experimental data. The transfer-matrix spectra are tested against those obtained using a full-wave, finite-element solver (COMSOL) that takes into account the exact form of eq. 4, finding nearly identical results. We note that in more complex structures, i.e., isolated or coupled nanoparticles, the approximation in eq. 12 may fail, and the complete version in eq. 4 should be used instead. The parameters of the nonlocal model of eq. 4 are retrieved according to the following procedure. First, we fix the background dielectric constant at $\varepsilon_\infty = 5.5$, in agreement with previous experimental investigations of the optical properties of doped CdO nanofilms. We then use the experimental values of the resonance wavelengths, $\lambda_m$, and the dispersion relation $\omega_m = \sqrt{\omega_p^2/\varepsilon_\infty + \frac{3}{5}v_F^2 q_m^2}$ in order to retrieve the values of $m^*$ and $n_0$. For the 20-nm-thick sample we find $m^* = 0.207\, m_e$ and $n_0 = 2.704 \cdot 10^{20}$ cm$^{-3}$, where $m_e$ is the electron rest mass. This value of carrier density is consistent with the doping conditions of the sputtering process, while the effective mass value is in line with that reported in[27]. A local scattering rate of $\gamma = 3.24 \cdot 10^{13}$ rad/s is deduced by the linewidth of the main resonance centered at $\lambda_1$, while a nonlocal scattering rate $\gamma_{VE} = 10\gamma$ is inferred by fitting the experimental spectra around the ancillary resonances at $\lambda_3$ and $\lambda_5$. In the simulations, the dielectric constants of MgO and CaF$_2$ are assumed dispersion-less and equal to 2.91 and 2.027, respectively. With these parameters, the nonlocal viscoelastic model is in excellent agreement with the experimental reflectivity data, as shown in Fig. 1. The only discrepancy between experiment and theory is found in the air-coupling case at short wavelengths: we attribute the difference to the angular sensitivity of $R_p/R_s$ around 45°, and some uncertainty in the precise value of the dielectric constant of MgO.

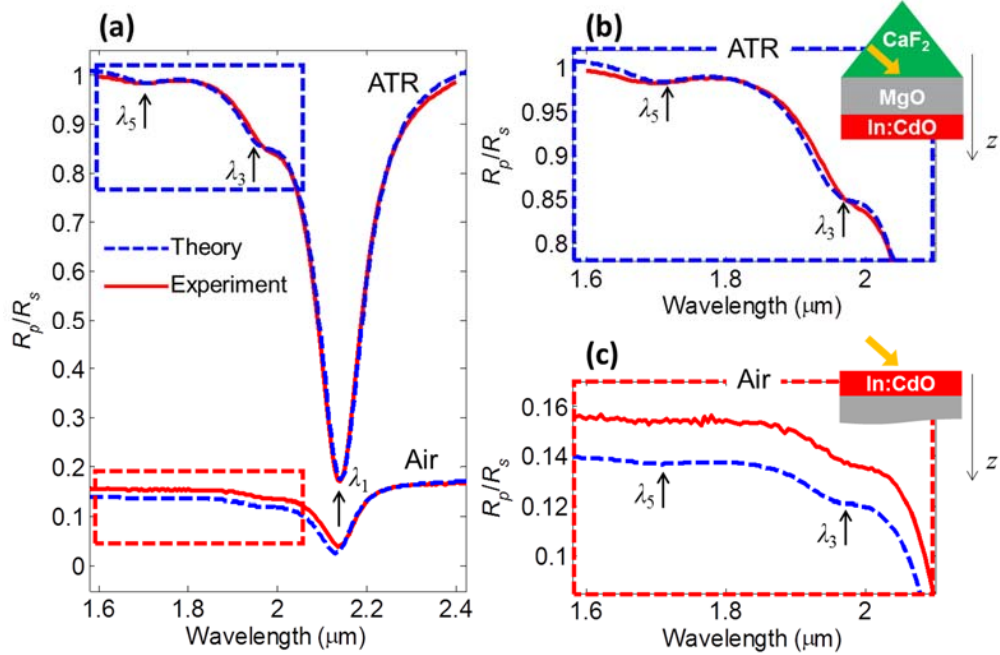

Fig. 1. (a) Normalized reflectivity spectra, $R_p/R_s$, for a 20-nm In:CdO film illuminated in the ATR and air-coupling configurations. (b) and (c) are zooms of (a) around the ancillary resonances at $\lambda_3$ and $\lambda_5$.

We stress that any attempt to fit the experimental spectra in Fig. 1 with a local permittivity model cannot reproduce the ancillary resonances at $\lambda_3$ and $\lambda_5$: these resonances can be explained only within the ambit of the nonlocal model of conduction electrons in the doped CdO film. We further verified that the observed optical behavior cannot be explained with the presence of few-nanometer-thick accumulation layers. Indeed, when such layers are properly modeled by solving Poisson's equation, the resulting smooth carrier density profile near the interface cannot explain any of the nonlocal effects observed here, namely the appearance of the ancillary resonances at $\lambda_3$ and $\lambda_5$, and the blueshift of the main resonance at $\lambda_1$ (this aspect will be discussed below). Moreover, the characterization of our In:CdO films, based on x-ray reflectivity and Hall-effect measurements, indicates the presence of depletion rather than accumulation layers, which have a negligible effect on the overall optical properties. In Fig. 2a we plot the dispersion relation of the longitudinal plasmon for the 20-nm film in the $\lambda$-$q$ plane, and highlight the positions of the $\lambda_m$ modes excited at odd integers of ($\pi/d$) and their near-perfect correspondence to the resonances observed in the experiment. For clarity, the reflection spectrum in Fig. 1a is also plotted in the limit of zero viscosity ($\gamma_{\text{VE}} = 0$), so that the dips are more visible.

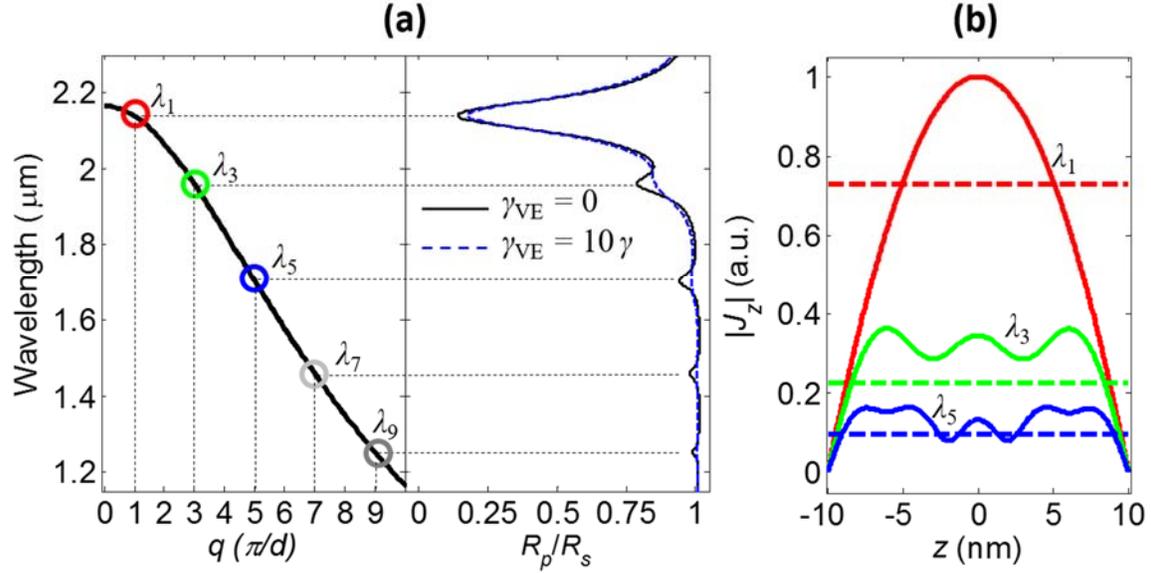

Fig. 2. (a) On the left, the dispersion of the longitudinal plasmon for the 20-nm In:CdO film – the longitudinal wavenumber $q$ on the x axis is in units of $\pi/d$; on the right, the reflectivity spectrum of the film in the presence ($\gamma_{VE}$ = 10 $\gamma$) and absence ($\gamma_{VE}$ = 0) of viscosity. (b) Amplitude of the longitudinal current $J_z$ within the thin film at the wavelengths of the first three resonances ($\lambda_1$, $\lambda_3$, and $\lambda_5$), calculated in the local approximation (dashed lines) and with the viscoelastic model using $\gamma_{VE}$ = 10 $\gamma$.

The well-known effect of the nonlocality is thus to introduce longitudinal electron pressure waves that resonates within the film at specific wavelengths, according to the dispersion relation $\omega(q)$. The interaction of these modes with light is not limited to the spectral features observed in Fig. 1 and 2a, but also involves the field distribution inside the film. In Fig. 2 we report the longitudinal current density, $|J_z|$, for the first three $\lambda_m$ modes, which tracks the fields inside the layer. The effect of the electron pressure is clearly recognizable as an oscillation of the induced current in the $z$-direction, with a number of peaks equal to the order number $m$. This is in stark contrast with the local approximation ($\mathbf{J}_{NL}=\mathbf{0}$), where the current density and the field are flat inside the film.

Another important aspect of the nonlocality is the thickness-dependence of the ENZ-modes' dispersion when $q \neq 0$. This leads to thickness-dependent dielectric constant and optical response. The effect is appreciable in Fig. 3, where we compare local and nonlocal absorption spectra calculated for In:CdO films with thicknesses varying from 5 to 30 nm in the ATR configuration at 45°. The viscoelastic parameters used in the nonlocal simulations are kept constant and equal to those extracted from the experimental data related to the 20-nm film.

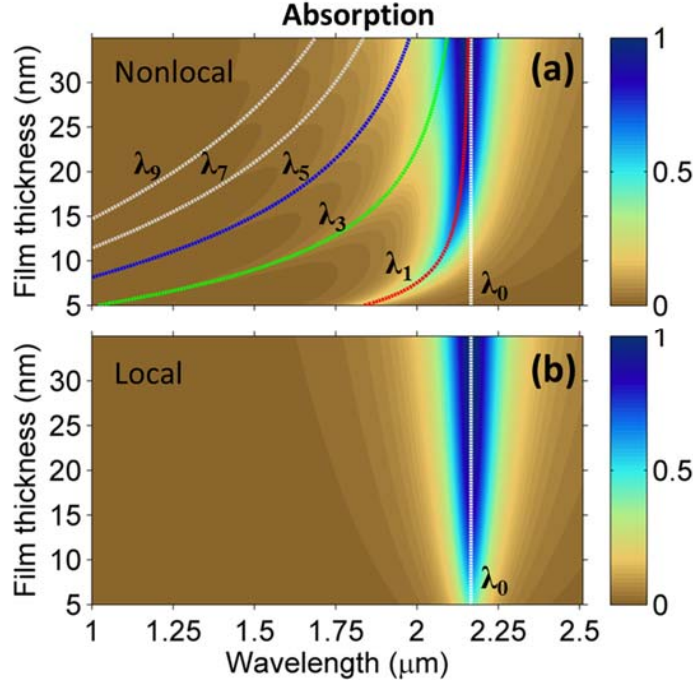

Fig. 3. (a) Color map plot of the absorption spectra (for *p*-polarization) of In:CdO films with thickness in the range 5-30 nm. The calculation is performed in the ATR configuration at 45° by adopting the viscoelastic parameters retrieved from the experimental spectrum of the 20-nm film. The overlapped curves correspond to the longitudinal plasmon modes wavelengths $\lambda_m$, with $m=1,3,5,\ldots$ The ENZ wavelength $\lambda_0 = 2\pi c \sqrt{\varepsilon_\infty}/\omega_p$ is plotted as a vertical line for reference. (b) Same as (a), but in the local approximation, i.e., $\mathbf{J}_{NL}=\mathbf{0}$.

Similarly to what may be observed in the reflection spectrum of Fig. 1, nonlocal absorption is characterized by a strong peak (close to 100%) corresponding to the first longitudinal mode at $\lambda_1$, whose dispersion overlaps the maximum absorption peak; shallower peaks are visible at the ancillary longitudinal plasmon modes: $\lambda_m = 2\pi c / \sqrt{\omega_p^2/\varepsilon_\infty + 3v_F^2 q_m^2/5}$. On the other hand, the local absorption spectrum shows a single well-defined peak at the fixed wavelength $\lambda_0 = 2\pi c \sqrt{\varepsilon_\infty}/\omega_p$, where the real part of the local (bulk) dielectric constant of In:CdO vanishes. As expected, the effect of the nonlocality is more discernable for thinner films, for which the wavelength of the main longitudinal plasmon $\lambda_1$ deviates more appreciably from the local prediction at $\lambda_0$. For example, for the 20-nm film of the experiment in Fig. 1, our model predicts a 25-nm blue-shift of the nonlocal peak with respect to the local one. A second-order approximation of the local-nonlocal deviation $\Delta\lambda = \lambda_0 - \lambda_1$ is given by

$$\Delta\lambda \approx \lambda_0 \frac{\varepsilon_\infty^2}{2\omega_p^2}\left(\sqrt{\frac{3}{5}}\frac{v_F \pi}{d}\right)^2. \tag{13}$$

For a 10 nm-thick film, the predicted blue-shift $\Delta\lambda$ is larger than 100 nm, according to eq. 13 and the results in Fig. 3. We have verified this behavior by fabricating a second In:CdO sample with approximately half the thickness of the sample discussed in Fig. 1. Using the same retrieval procedure outlined for the 20-nm film, we fixed $\varepsilon_\infty = 5.5$ as in[27] and extracted the following viscoelastic parameters: $m^* = 0.230\ m_e$, $n_0 = 3.223 \cdot 10^{20}$ cm$^{-3}$, $\gamma = 2.513 \cdot 10^{13}$ rad/s, $\gamma_{VE} = 15\gamma$ and thickness $d = 11$ nm. The fact that the effective mass in this sample is slightly larger than in the 20-nm sample is most likely due to the larger carrier concentration. Indeed, this effective-mass dependence on the carrier concentration has been previously observed in In:CdO and attributed to the non-parabolicity of the conduction band[47-48]. Moreover, these discrepancies of effective mass, carrier density and scattering rates in this sample with respect to those found in the 20-nm sample to the unavoidably different sputtering conditions under which the two samples were fabricated. The reflection spectra of the film are reported in Fig. 4 for both the ATR and the air-coupling configurations, showing excellent agreement of the viscoelastic theory with the experiment in both cases.

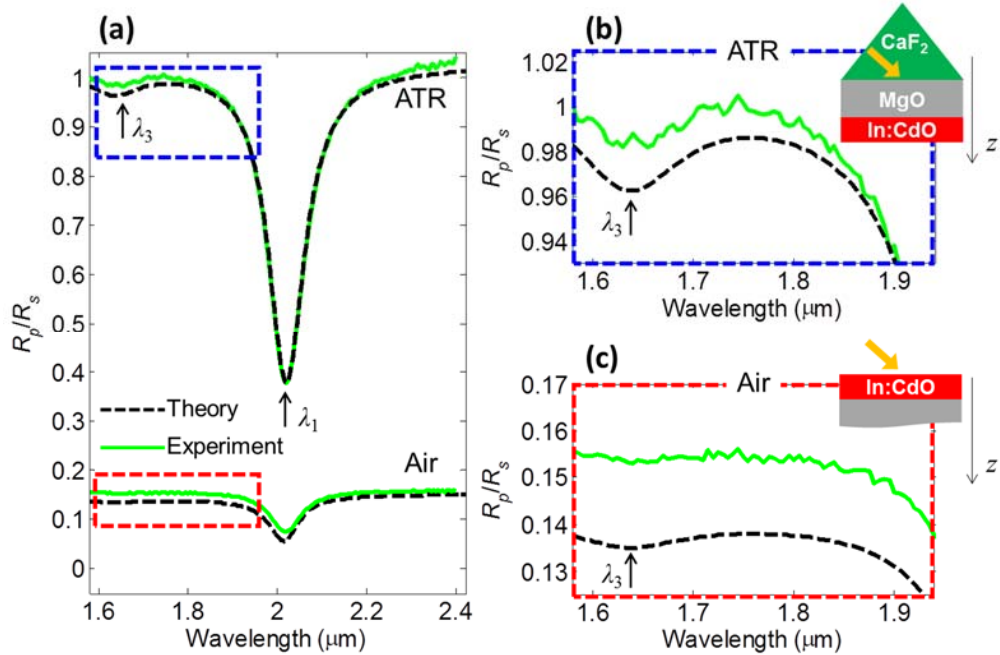

Fig. 4. (a) Normalized reflectivity spectra, $R_p/R_s$, for an 11-nm In:CdO film illuminated in the ATR and air-coupling configurations. (b) and (c) are zooms of (a) around the ancillary resonances at $\lambda_3$ and $\lambda_5$.

The fundamental mode is near $\lambda_1 = 2.02$ μm, blue-shifted by approximately 70 nm with respect to the bulk ENZ wavelength $\lambda_0$ and in line with the shift prediction $\Delta\lambda$ based on eq. 13. The only ancillary resonance discernible in this wavelength range is the one associated with the mode of order three at $\lambda_3 \sim 1.6$ μm. The slight discrepancies between experiment and theory in the air-coupling configuration are similar in nature to those found in the case of the 20-nm film.

*Damping and viscosity*

The ENZ film has two sources of damping: the local damping is related to the real part of the conductivity $\sigma$, or to the damping constant $\gamma$ appearing in the (local) Drude model; the nonlocal damping is due to viscosity of the electron fluid moving under the influence of an ionic background, and is accounted for by the terms proportional to $\gamma_V \nabla(\nabla \cdot \mathbf{J})$ – see eq . 4. A simple way to appreciate the effects of these damping sources in our ENZ film is to calculate the loss function, $\mathrm{Im}(-\varepsilon_\ell^{-1})$, a quantity related to the rate at which charged particles lose energy in a medium[49], and that can be directly measured with electron energy loss spectroscopy. In the ENZ regime, close to the zero-crossing frequency of the permittivity's real part, the loss function shows a peak approximately equal to $1/\mathrm{Im}(\varepsilon_\ell)$, suggesting that absorption losses decrease as damping increases, as recently discussed in[50]. From the expression of $\varepsilon_\ell$, it is clear that viscosity alters the dielectric permittivity by introducing additional (nonlocal) damping. This effect is visible in the map of the In:CdO loss function, plotted in Fig. 5(a) as a function of wavelength and longitudinal wavenumber. The maximum peak of the loss function is obtained in the local limit, i.e., when $q = 0$ and the damping is equal to $\gamma$. In the nonlocal regime, the loss function peak decreases and broadens with increasing longitudinal wavenumbers, as a result of increasing viscosity and damping. The importance of viscosity can then be appreciated when one compares the loss functions in the presence vs absence of viscosity, as described in Fig. 5b for a wavenumber $q = 1$ nm$^{-1}$. The peak in the viscous case is quite shallower and lower in intensity, indicating a decreased ability to absorb photon (or electron) energy. At the same time, far from the ENZ regime, the loss function (hence absorption losses) in the viscous case is larger than the corresponding non-viscous case. An important effect related to viscosity is that nonlocal damping leads to thickness-dependent damping. Let us consider the linewidth of the main ENZ mode at $\lambda_1$. Based on the longitudinal relative permittivity in eq. 11, we can write a second-order approximated expression of the total damping in the film at the wavelength $\lambda_1$ as

$$\gamma_{\mathrm{NL}} \approx \gamma \left[1 + \frac{4\varepsilon_\infty \alpha_{\mathrm{VE}} \pi^2 v_F^2}{15 \omega_p^2 d^2}\right], \tag{14}$$

where the fitting factor $\alpha_{\mathrm{VE}} = \gamma_{VE}/\gamma$ was found to be equal to 10 in the case of the 20-nm film, and 15 in the case of the 11 nm-thick film. According to eq. (14), total damping increases as film thickness decreases, a well-known effect[51] that can also be explained in terms of diffusion[42]. This behavior can be clearly recognized in Fig. 3a, where the absorption resonance at $\lambda_1$ is increasingly damped as the thickness decreases, in response to the increasing value of $\gamma_{\mathrm{NL}}$ and decreasing loss function. Besides damping and viscosity, we point out that spectral broadening of plasmonic resonances may also be due to surface roughness in our samples. We have verified through numerical simulations that the introduction of a roughness of approximately 1 nm barely affects the main resonance of the film, and slightly broadens the ancillary resonances, without altering their spectral position.

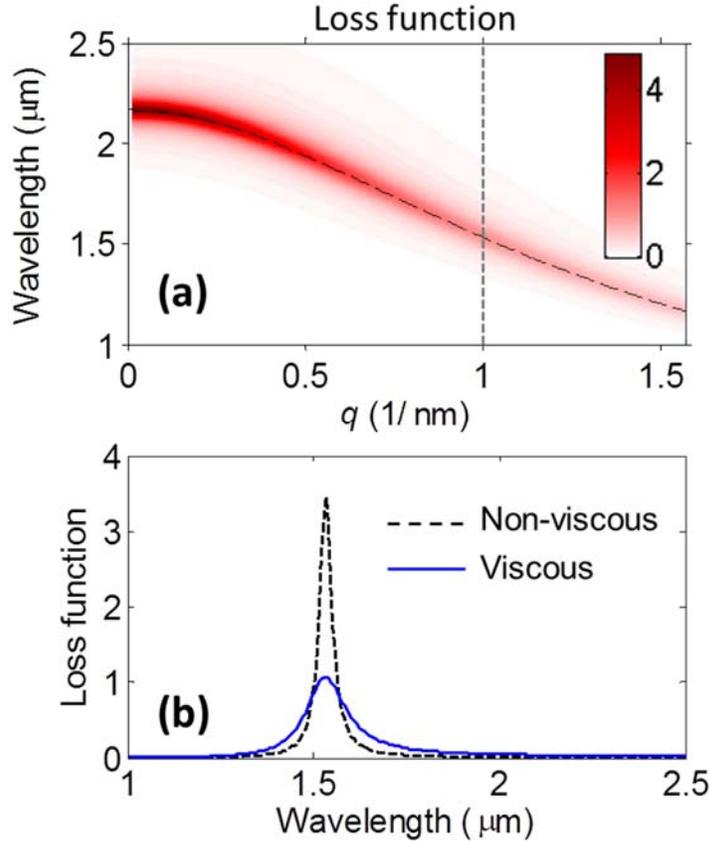

Fig. 5. (a) Color map of the loss function, $\mathrm{Im}(-\varepsilon_\ell^{-1})$, as a function of wavelength and longitudinal wavenumber $q$. The dashed-black curve is the dispersion of the longitudinal plasmon $\lambda(q) = 2\pi c / \omega(q)$ where $\omega(q) = \sqrt{\omega_p^2/\varepsilon_\infty + \frac{3}{5}v_F^2 q^2}$. (b) Loss function spectrum for $q = 1$ nm$^{-1}$ in the viscous [blue curve, i.e., cross-section of the map in (a) at the position marked by the vertical line] and non-viscous (dashed-black curve) cases.

## Conclusions

In summary, the nonlocal optical response of indium-doped CdO nanofilms has been experimentally observed and quantified using optical spectroscopy. A hydrodynamic, viscoelastic theory for the motion of conduction electrons has been used to reproduce the experimental findings. The elastic behavior of conduction electrons produces a significant blue-shift of ENZ modes. The main film resonance, associated with the longitudinal plasmon of order 1, undergoes a blue-shift of approximately 25 nm for a 20-nm-thick film, and 70 nm for an 11-nm-thick film, as predicted by our viscoelastic nonlocal theory. Shallower, secondary resonances associated with higher-order longitudinal plasmons at frequencies larger than the screened (bulk) plasma frequency can be clearly observed as dips in reflection and peaks in absorption spectra. We find that the electron-fluid viscosity plays an important, non-trivial role in the optical response, by introducing thickness-dependent and spatially-dispersive (nonlocal) damping. Differently from previous treatments , the viscosity in our model is self-consistently derived from the trace-less part of the stress tensor rather than being introduced phenomenologically as a diffusion current[42] or as a correction term[44].

## Acknowledgments

This work was supported by the US Department of Energy (DOE), Office of Basic Energy Sciences, Division of Materials Sciences and Engineering and performed, in part, at the Center for Integrated Nanotechnologies, an Office of Science User Facility operated for the US DOE Office of Science, and in part supported by the Laboratory Directed Research and Development program at Sandia National Laboratories. Sandia National Laboratories is a multimission laboratory managed and operated by National Technology and Engineering Solutions of Sandia, LLC., a wholly owned subsidiary of Honeywell International, Inc., for the U.S. Department of Energy's National Nuclear Security Administration under contract DE-NA-0003525. K.K. and J-P.M. were supported by NSF grant CHE-150794 and ARO grant W911NF-16-1-0037. D.d.C. thanks Cristian Ciracì from Istituto Italiano di Tecnologia for helpful discussions.

## Authors contributions

D.d.C, M.S. and M.A.V developed the theory, performed numerical simulations and wrote the initial draft of the manuscript. S.C., G.A.K., T.S.L. designed and performed the optical characterization. K.K., E.L.R and J-P.M. prepared the samples. All the authors discussed the results, contributed to the writing and reviewed the manuscript.

## Competing interests

The authors declare no competing financial interests.

## Data Availability

The datasets generated during and/or analyzed during the current study are available from the corresponding author on reasonable request.